\documentclass[10pt,conference]{IEEEtran}
\usepackage{cite}
\usepackage{amsmath,amssymb,amsfonts}
\usepackage{algorithmicx}
\usepackage{graphicx}
\usepackage{textcomp}
\usepackage{xcolor}
\usepackage[hyphens]{url}
\usepackage{fancyhdr}
\usepackage{subcaption}
\usepackage{svg}
\usepackage{algorithm}
\usepackage{algpseudocode}
\usepackage{tikz}
\usetikzlibrary{quantikz2-akj}
\usepackage{xcolor}
\usepackage{ifthen}
\usepackage{mathabx}
\newboolean{showcomments}
\setboolean{showcomments}{true}
\ifthenelse{\boolean{showcomments}}
{ \newcommand{\mynote}[3]{
     \fbox{\bfseries\sffamily\scriptsize#1}
        {\small$\blacktriangleright$\textsf{\emph{\color{#3}{#2}}}$\blacktriangleleft$}}}
{ \newcommand{\mynote}[3]{}}
\definecolor{asparagus}{rgb}{0.53, 0.66, 0.42}

\newcommand{\cnot}{\texttt{CNOT}}

\newcommand{\sqiswap}{$\sqrt{\texttt{iSWAP}}$}
\newcommand{\niswap}[1]{$\sqrt[#1]{\text{\texttt{iSWAP}}}$}
\newcommand{\iswap}{\texttt{iSWAP}}
\newcommand{\sw}{\texttt{SWAP}}

\newcommand{\pswap}{\texttt{pSWAP}}

\newcommand{\cns}{\texttt{CNS}}

\newcommand{\cphase}{\texttt{CPHASE}}
\usepackage[hyphens]{url}
\usepackage{hyperref}

\title{MIRAGE: Quantum Circuit Decomposition and Routing Collaborative Design using Mirror Gates}
\def\hpcacameraready{} % Uncomment to build camera-ready version

\newcommand\hpcaauthors{Evan McKinney$\dagger$, Michael Hatridge$\ddagger$, Alex K. Jones$\dagger$}
\newcommand\hpcaaffiliation{Department of Electrical and Computer Engineering$\dagger$, Department of Physics and Astronomy$\ddagger$,\\ University of Pittsburgh}
\newcommand\hpcaemail{evm33@pitt.edu, hatridge@pitt.edu, akjones@pitt.edu}

\author{
  \ifdefined\hpcacameraready
    \IEEEauthorblockN{\hpcaauthors{}}
      \IEEEauthorblockA{
        \hpcaaffiliation{} \\
        \hpcaemail{}
      }
  \else
    \IEEEauthorblockN{\normalsize{HPCA \hpcayear{} Submission
      \textbf{\#\hpcasubmissionnumber{}}} \\
      \IEEEauthorblockA{
        Confidential Draft \\
        Do NOT Distribute!!
      }
    }
  \fi 
}

\fancypagestyle{camerareadyfirstpage}{%
  \fancyhead{}
  
  \fancyhead[C]{} % clear the header content here
  \fancyfoot[C]{}
}

\begin{document}
\maketitle

%Enables the camera ready header and footer
\ifdefined\hpcacameraready 
  \thispagestyle{camerareadyfirstpage}
  \pagestyle{empty}
\else
  \thispagestyle{plain}
  \pagestyle{plain}
\fi

\newcommand{\hpcaheight}{0mm}
\ifdefined\eaopen
\renewcommand{\hpcaheight}{12mm}
\fi

\thispagestyle{plain} % enables page number on the first page
\pagestyle{plain} % enables page numbers on the rest of the document

%%%%%%%%%%%%%%%%%%%%%%%%%%%%%%%%%%%%%%%%
%%%%%%%% -- PAPER CONTENT STARTS -- %%%%%%%%%

\begin{abstract}
Building efficient large-scale quantum computers is a significant challenge due to limited qubit connectivities and noisy hardware operations. Transpilation is critical to ensure that quantum gates are on physically linked qubits, while minimizing $\texttt{SWAP}$ gates and simultaneously finding efficient decomposition into native $\textit{basis gates}$. The goal of this multifaceted optimization step is typically to minimize circuit depth and to achieve the best possible execution fidelity. In this work, we propose $\textit{MIRAGE}$, a collaborative design and transpilation approach to minimize $\texttt{SWAP}$ gates while improving decomposition using $\textit{mirror gates}$. Mirror gates utilize the same underlying physical interactions, but when their outputs are reversed, they realize a different or $\textit{mirrored}$ quantum operation. Given the recent attention to $\sqrt{\texttt{iSWAP}}$ as a powerful basis gate with decomposition advantages over $\texttt{CNOT}$, we show how systems that implement the $\texttt{iSWAP}$ family of gates can benefit from mirror gates. Further, $\textit{MIRAGE}$ uses mirror gates to reduce routing pressure and reduce true circuit depth instead of just minimizing $\texttt{SWAP}$s. We explore the benefits of decomposition for $\sqrt{\texttt{iSWAP}}$ and $\sqrt[4]{\texttt{iSWAP}}$ using mirror gates, including both expanding Haar coverage and conducting a detailed fault rate analysis trading off circuit depth against approximate gate decomposition. We also describe a novel greedy approach accepting mirror substitution at different aggression levels within MIRAGE. Finally, for $\texttt{iSWAP}$ systems that use square-lattice topologies, $\textit{MIRAGE}$ provides an average of 29.6\% reduction in circuit depth by eliminating an average of 59.9\% $\texttt{SWAP}$ gates, which ultimately improves the practical applicability of our algorithm.
\end{abstract}
% We also describe two novel approaches: first, a greedy approach accepting mirror substitution at different aggression levels; second, a simulated annealing inspired approach improves on the greedy technique.  

\section{Introduction}
\label{sec:introduction}
% NISQ solves problems - but has challenges
Quantum computers attempt to leverage superposition states and entanglement between multiple quantum bits or \textit{qubits} to efficiently solve problems that are intractable for classical computers. These operations between qubits form quantum \textit{gates} that collectively form quantum \textit{circuits}.  However, current ``noisy, intermediate scale quantum'' (NISQ) machines face limited connectivity between qubits and significant reliability challenges from executing these quantum circuits. These challenges are from many sources of noise including energy decay, dephasing, and crosstalk among qubits, etc~\cite{nielsen2002quantum}. Thus, the holistic co-design goal for the execution of algorithms on quantum machines is to minimize the depth of the circuit, since the depth of the circuit is directly related to the fidelity of the circuit through the cumulative effect of these internal and external sources of noise~\cite{emerson2005scalable, gao2021practical}.  %These factors and circuit fidelity can be related to the depth of the underlying quantum circuit to be executed on the quantum machine.

% Factors that impact circuit depth circuit decomposition and data movement on the target quantum machine. 
Different quantum machines use different \emph{basis gates} governed, in part, by the type of physical interactions inherent to the quantum hardware.  In superconducting NISQ machines, IBM uses the cross-resonance basis gate, a gate that is locally equivalent to a \cnot{}~\cite{rigetti2010fully}.  This choice is convenient since many quantum algorithms are themselves written in the \cnot{} basis.  Other basis gates include the Google \texttt{SYC} gate and the \iswap{} gate, which is readily realizable by systems using Transmon qubits~\cite{arute2019quantum, sung2021realization, roth2017analysis}.  The number of basis operations required is determined by circuit decomposition.  Moreover, restricted quantum topologies require the use of \sw{} gates to move information between qubits% as direct connections between qubits are often not directly implementable
.  \sw{} gates must also be decomposed into %the quantum computer 
basis gates, and generally require the most basis gate operations to be realized.

% Define the purpose of transpiler
% efficient decomposition reduces gates (we go on to explore using mirror gates and approximation), and even more important is reducing SWAPs (we go on to explore efficient routing algorithms
The process to decompose, place gates, and route (i.e., insert \sw{} gates) on the physical machine is called transpilation.  Minimizing circuit depth in the transpiler is NP-hard~\cite{cowtan2019qubit}, thus, state-of-the-art techniques are largely heuristic, ranging from stochastic methods to subgraph isomorphism algorithms~\cite{nannicini2022optimal}. Due to the complexity, a customary abstraction barrier is inserted between routing and decomposition such that they are computed independently.  This is designed to make the problems independent and tractable.

Unfortunately, this separation of concerns often disregards scenarios where a certain choice of where a \sw{} is inserted can enhance quantum circuit compression, an advantage only detectable post basis-translation~\cite{liu2023qcontext}. One significant example includes a two-qubit gate $U$ immediately followed by a \sw{} on the same circuit wires, which together form $U'$, denoted as the \textit{mirror gate} of $U$~\cite{jurcevic2021demonstration, cross2019validating}.

% \begin{figure}[tbp]
%     \centering
%     \begin{subfigure}[b]{\columnwidth}
%         \includegraphics[width=\columnwidth]{cnot_decomposition.pdf}
%         \caption{\cnot{} decomposes into two \sqiswap{} gates}
%         \label{fig:cnot_decomposition}
%     \end{subfigure}
%     \begin{subfigure}[b]{\columnwidth}
%         \includegraphics[width=\columnwidth]{cns_decomposition.pdf}
%         \caption{\cnot{}+\sw{} (\cns{}) also decomposes into 2 \sqiswap{} gates}
%         \label{fig:cns_decomposition}
%     \end{subfigure}
%     \caption{Decomposition of \cnot{}, \cns{} into the \iswap{} family}
%     \label{fig:gate_decompositions}
% \end{figure}

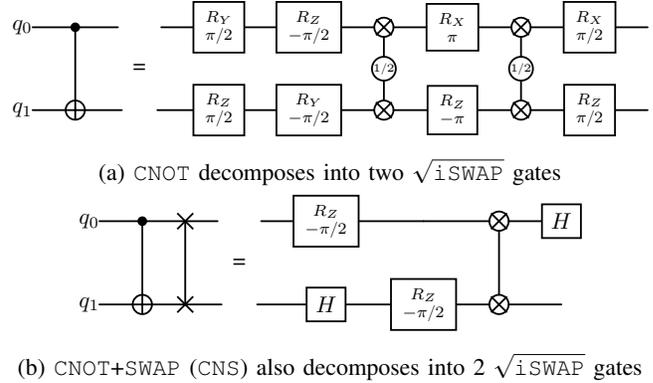
\begin{figure}[tbp]
    \centering
    \begin{subfigure}[b]{\columnwidth}
    \resizebox{\columnwidth}{!} {
    \begin{quantikz}
    q_0 & \ctrl{1} & \midstick[2,brackets=none]{=}  &\gate{R_Y \atop \pi/2}&\gate{R_Z \atop -\pi/2}& %\iSwap[partial swap={\frac{1}{2}}]{1} 
    \iSwap[partial swap=1/2]{1}
    & \gate{R_X \atop \pi} & 
    %\iSwap[partial swap={\frac{1}{2}}]{1} 
    \iSwap[partial swap=1/2]{1}
    & \gate{R_X \atop \pi/2} & 
    \\
    q_1 & \targ{} &  &  \gate{R_Z \atop \pi/2}&\gate{R_Y \atop -\pi/2}& \targiS{} &\gate{R_Z \atop -\pi}&\targiS{}&\gate{R_Z \atop \pi/2} &
    \end{quantikz}
    }
        \centering
        \caption{\cnot{} decomposes into two \sqiswap{} gates}
        \label{fig:cnot_decomposition}
    \end{subfigure}
    
    \begin{subfigure}[b]{\columnwidth}
        \centering
    \resizebox{.8\columnwidth}{!} {
    \begin{quantikz}
    q_0 & \ctrl{1} & \swap{1} & \midstick[2,brackets=none]{=} & 
    \gate{R_Z \atop -\pi/2} & & \iSwap{1} & \gate{H}
    \\
    q_1 & \targ{} & \targX{} & & \gate{H} & 
    \gate{R_Z \atop -\pi/2} & \targiS{} & 
    \end{quantikz}
    }
        \caption{\cnot{}+\sw{} (\cns{}) also decomposes into 2 \sqiswap{} gates}
        \label{fig:cns_decomposition}
    \end{subfigure}
    \caption{Decomposition of \cnot{}, \cns{} into the \iswap{} family}
    \label{fig:gate_decompositions}
\end{figure}

A known mirror gate example is the relationship of \cnot{} and \iswap{}, shown in Fig.~\ref{fig:gate_decompositions}.  \cnot{} appears ubiquitously in quantum circuits due to its ability to create entangled states as well as defining subroutines including singular value transformations, stabilizer measurements, and the decomposition of multi-qubit gates~\cite{nielsen2002quantum}.  Of course, as stated above, many superconducting qubits natively produce photon-exchange gates like \iswap{}~\cite{zhou2023realizing, sung2021realization, mundada2019suppression}, or, in bosonic cases, `beam-splitters'~\cite{robstuff}.  Moreover, recent work showed that \cnot{} could be decomposed into two \sqiswap{} gates (Fig.~\ref{fig:cnot_decomposition})~\cite{huang2023quantum}.  However, the \cnot{} mirror gate, \cns{} or $\cnot{}+\sw{}$, is locally equivalent to an \iswap{}, i.e., two \sqiswap{}s (Fig.~\ref{fig:cns_decomposition})~\cite{schuch2003natural}. Thus, in the \sqiswap{} basis, \cnot{} and \cns{} have the same circuit depth cost.
%—to further explore the utility of mirror gates. 
\textit{This ``free'' data movement of \cns{} using \iswap{} can be leveraged for quantum circuit depth compression during transpilation.}

Thus, we propose MIRAGE or \textit{Mirror-decomposition Integrated Routing for Algorithm Gate Efficiency}.  MIRAGE is a quantum co-design methodology based on utilizing mirror gates to aid in both decomposition and routing quantum circuits onto quantum machines. In MIRAGE we consider a \textit{mirage} \sw{} to be a \sw{} gate that can be absorbed into another computational gate during decomposition, as in a \cns{} decomposing into \iswap{} as in Fig.~\ref{fig:cns_decomposition}.%  Thus, this \textit{mirage} gate is like a mirage because it disappears in the decomposed circuit. \mjh{you can ditch this last sentence, too obvious}

% MIRAGE also has the goal to use this mirage \sw{} gate to improve the circuit routing.  
% We propose MIRAGE to create a transpilation algorithm that attempts to leverage mirror gates to improve routing.  
% ^^ getting too repetetive proposing MIRAGE here
Moreover, MIRAGE, unlike SABRE, breaks down the barrier between routing and decomposition.  MIRAGE considers both the routing impact routing (adding \sw{} gates) and the decomposition depth when selecting between a gate or its mirror.  %In developing MIRAGE, due to its heuristic nature, SABRE is highly impacted by starting choices for qubit placement.  
Like other heuristics, in MIRAGE starting conditions and condition(s) when to select mirror gates are highly impactful.  Thus, we explore different \textit{aggression levels}, which use different thresholds on when to insert mirror gates.  %We also compare this to a simulated annealing inspired version of MIRAGE in an attempt to be less impacted by initial conditions and thresholds as a mechanism to insert mirror gates.

%The  The \cnot{} and \sw{} gates together form a unitary that is locally equivalent to an \iswap{}~\cite{schuch2003natural} as illustrated in Fig.~\ref{fig:gate_decompositions}.  Fig.~\ref{fig:cnot_decomposition} shows the decomposition of \cnot{} into two \sqiswap{} gates, and Fig.~ \ref{fig:cns_decomposition} shows the decomposition of \cnot{}+\sw{} (the mirror of \cnot{} sometimes called `\cns{}`) into two \sqiswap{} gates. \textbf{The key observation is that the cost of \cnot{}, and the mirror of \cnot{} are equivalent in the \sqiswap{} basis.} This ``free'' data movement baked into the \cns{} operation can be leveraged for depth compression during transpilation.

% To address these key challenges of executing high-fidelity quantum algorithms, we research advancements in efficient transpilation.
% \mjh{something broken here, not sure what's going on}
% % list contributions
% First, we delve into how mirror gates can facilitate decreased expected costs using approximate decompositions. Second, we develop an optimized routing algorithm that factors in the cost of mirror gates for more efficient data movement. This flexible routing strategy adapts to the unique characteristics of each quantum circuit, reducing key sources of error and boosting overall circuit fidelity. In particular:
%\evm{Fix placeholders}

In particular, we propose the following contributions:
\begin{enumerate}
    \item We demonstrate and quantify the value of using mirror gates for the \iswap{} family of basis gates (i.e., \sqiswap{}, \niswap{3}, and \niswap{4}) basis gates. 
    \item We demonstrate mirror gates provide a similar decomposition benefit to approximate decomposition and that both approaches can be combined for additional reduction of infidelity and improved Haar scores. %When combined, mirror gates and approximate decomposition  provide nearly 9\% infidelity reduction and Haar score benefits to circuits mapped to a \sqiswap{} basis.
    \item We develop the MIRAGE transpilation flow that leverages mirror gates to benefit routing and decomposition.
    \item %Modify selection over independent trials; 
    We show when specifically optimizing for circuit depth rather than minimizing \sw{} gates can achieve a significant improvement of 7.5\%.
    \item We propose aggression levels within the MIRAGE framework demonstrate effective transpilation of different classes of circuits. %On the Heavy-Hex topology, we achieved an average depth reduction of 31.19\%. On the Square-Lattice topology, our optimizations led to an average decrease of 29.58\% circuit depth.
    % \item We created a simulated annealing inspired version of MIRAGE which is less invariant to starting conditions and thresholds used by MIRAGE to determine when to insert mirror gates.
\end{enumerate}
% From our theoretical and experimental studies we determine that using mirror gates with gate approximation can provide nearly a 9\% reduction of infidelity of circuits with a \sqiswap{} basis gate. \akj{summarize transpiler results} %The results of our experiments demonstrate a reduction in depth and, in turn, will boost fidelity of circuit execution on near-term hardware.
Our results indicate that mirror gates with approximate decomposition can reduce infidelity by nearly 9\%.  Additionally, MIRAGE can further improve infidelity through approximately 30\% circuit depth reductions over Qiskit. 
 Before we describe the details of our \iswap{} family mirror gate analysis and the MIRAGE transpilation approach, we first provide some background in the next section.

\section{Background}
\label{sec:background}
Quantum transpilers \raisebox{.5pt}{\textcircled{\raisebox{-.9pt} {1}}}~decompose gates in quantum algorithms onto the basis gate realizable in the target system, \raisebox{.5pt}{\textcircled{\raisebox{-.9pt} {2}}}~physically place the qubits and gates in the mapped algorithm to locations on the target machine, and \raisebox{.5pt}{\textcircled{\raisebox{-.9pt} {3}}}~insert \sw{} gates into the circuit to route states between qubits. 

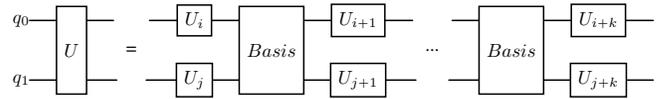
\begin{figure}[tbp]
    \resizebox{\columnwidth}{!} {
    \begin{quantikz}
    q_0 & \gate[2]{U} &  \midstick[2,brackets=none]{=} & \gate{U_i} & \gate[2]{Basis} & \gate{U_{i+1}} & \midstick[2,brackets=none]{...} & \gate[2]{Basis} & 
    \gate{U_{i+k}} &
    \\
    q_1 & & & \gate{U_j} & & \gate{U_{j+1}} & & & \gate{U_{j+k}} &
    \end{quantikz}
    }
    \caption{Arbitrary basis gate decomposition template}
    \label{fig:basis-gate-decomp}

\end{figure}
A common method to decompose an arbitrary two-qubit (2Q) gate, or \textit{unitary} $U$ in a basis gate uses Cartan's decomposition.  In Cartan's KAK decomposition
%, named for the mathematical expression, 
a 2Q $U$ is factored by a series of 2Q basis gates combined with single-qubit unitaries(1Q)~\cite{tucci2005introduction}. This method allows for the effective construction of any 2Q unitary operation as shown in Fig.~\ref{fig:basis-gate-decomp} for a generalized form that requires $k$ copies of the basis gate.

% This feels tangent and inconsequential - we've already said topologies require SWAPs a couple times
Placement and routing are important because quantum topologies are restricted. While many machines use a square lattice topology, as IBM machines scale, their routing flexibility eventually decreases to a heavy-hex lattice, which is currently standard~\cite{chow2011simple}.  This routing limitation is due to the inherent crosstalk between qubits of the IBM cross-resonance gate~\cite{yan2018tunable}.  Thus, as topologies become increasingly limited, the number of \sw{} gates required to implement circuits increases.

However, deep circuits work against circuit fidelity because infidelity is quantitatively linked to the total execution time of the quantum circuit. Specifically, the overall cost in time is derived from the durations of individual 1Q and 2Q gates, along with the number of necessary $k$ applications of the 2Q basis gate. As the duration, or depth, of a quantum circuit increases, the system becomes increasingly susceptible to noise, such as decoherence and dephasing, leading to a gradual degradation of the stored quantum information~\cite{unruh1995maintaining}. Additionally, the inclusion of \sw{} operations contributes significantly to the overall duration and, consequently, the infidelity of executing a quantum circuit. Notably, for \niswap{n} basis gates, \sw{} operations demand the highest $k$ values for decomposition, thereby becoming the dominant source of error~\cite{huang2023quantum, mckinney2023parallel}.

% Overview of previous transpilation optimization works
Various transpilation optimization techniques have been developed in recent years, such as finding gate commutation rules~\cite{bowman2022hardware}, pulse level decomposition optimizations~\cite{earnest2021pulse} and eliminating gates for known pure-state inputs~\cite{liu2021relaxed}. However, The effectiveness of these solutions are highly context dependent, emphasizing the importance of comparative evaluations in diverse quantum computing scenarios.

% TODO integrate this better with previous section. We want to give a little more rigor to the idea of CNS, and in general mirrors using our friend, the weyl chamber :)

In the next section we describe the impact of using mirror gates in decomposition.

\section{Mirror-Gate Decomposition}
\label{sec:mirror-gate-decomposition}
% in this section...
To understand the potential impact of decomposition using mirror gates requires a method for articulating how this process can contribute to circuit depth reduction.  In this section we compute the advantage on computational power of different basis gates using mirror gates on unrestricted topologies (e.g., all-to-all networks), and then we explore how this maps to different restricted topologies.  However, to define and understand certain terms, methodologies, and metrics, we start with some preliminaries discussed in the next section.

\subsection{Preliminaries}
\label{sec:mirror-gate-decomposition:sub:preliminaries}

%discussion of the potential impact
% Define exact decomposition as open problem
In Section~\ref{sec:background} we introduced the concept of decomposition into basis gates using Cartan's Decomposition.  However, decomposition into sequences only the target hardware's basis gates has been a study of considerable research that can be divided into two categories: exact analytical decomposition and approximate numerical decomposition. For 2Q unitary targets, an optimal explicit set of rules has been derived for \cnot{}~\cite{vatan2004optimal} and much more recently for the \sqiswap{}~\cite{huang2023quantum}. However, generalizing decomposition to an arbitrary-sized unitary target is an open problem with the current most efficient analytical method, Quantum Shannon Decomposition (QSD) still far from optimal. For instance, QSD decompositions to \cnot{} gates produces solutions approximately twice as expensive as the known theoretical lower bound~\cite{rakyta2022approaching}. Moreover, QSD and other methods such as Cosine-Sine Decomposition (CSD) only work for controlled-unitary basis gates \cite{saeedi2010block, shende2005synthesis}.
% redundant.
%, which do not work for other relevant basis gates, such as the \syc{} or the \iswap{} families of gates. 

% Numerical methods
In contrast, numerical decomposition methods are more flexible than analytical decomposition and can closely approach the theoretical lower bounds of gate costs, even when scaling up to 5Q decompositions~\cite{rakyta2022approaching, rakyta2022efficient, osti_1785933}. Numerical decomposition attempts to tune the parameters of a \textit{circuit ansatz}.  The ansatz is essentially a starting guess for the form of the decomposed circuit, such that the better the form of the ansatz matches the optimal decomposition, the better the quality of the numerical solution. 

As in Cartan's decomposition from Fig.~\ref{fig:basis-gate-decomp}, a reasonable numerical approach sequentially alternates applications of the 1Q and 2Q gates, to form different choices of the ansatz in an attempt to best match the form of an optimal or near-optimal solution. %not unlike from Cartan's KAK decomposition (Fig.~\ref{fig:basis-gate-decomp}). 
The suitability of the ansatz is determined through numerical optimization of the 1Q gates.  The suitability of the solution can be measured using a similarity approach such as the Hilbert-Schmidt norm, which defines the distance between the ansatz and the current unitary~\cite{lao2021designing}.  However, the use of numerical methods becomes complex with an increase in the number of qubits, due to an exponentially expanding parameter space and the intricate generation of the ansatz.

\begin{figure}[tbp]
    \centering
    \begin{subfigure}[b]{.45\columnwidth}
    \includegraphics[width=\linewidth]{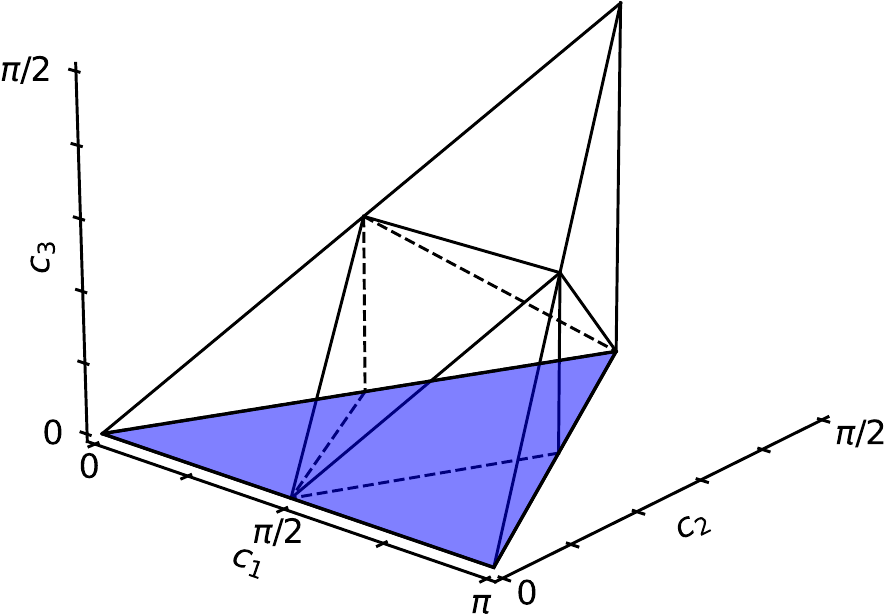}
    \caption{Polytope for \cnot{}}    
    \label{fig:poly-cnot}
    \end{subfigure}  
    \begin{subfigure}[b]{.45\columnwidth}
    \includegraphics[width=\linewidth]{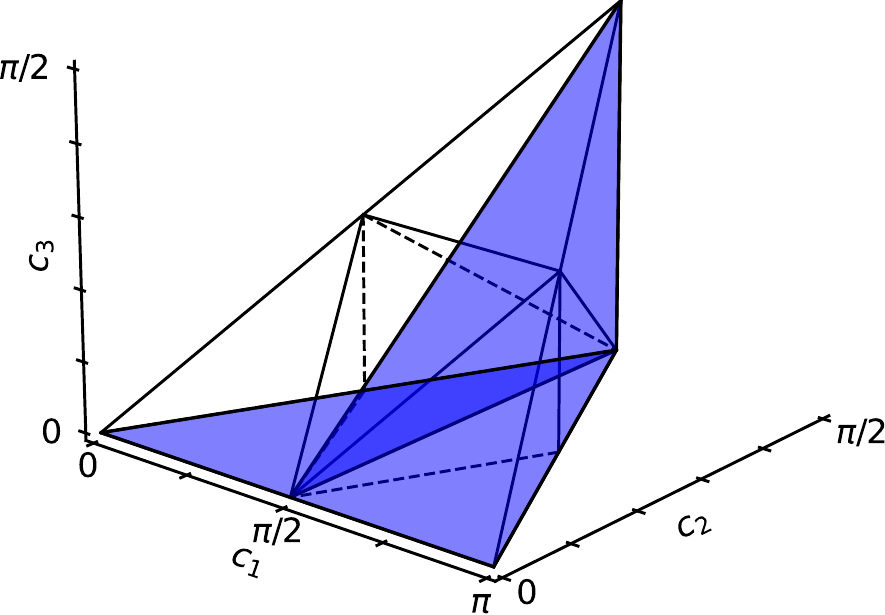}
    \caption{\cnot{} with mirror gates}    
    \label{fig:poly-cnot-mirror}
    \end{subfigure}        
    \begin{subfigure}[b]{.45\columnwidth}
    \includegraphics[width=\linewidth]{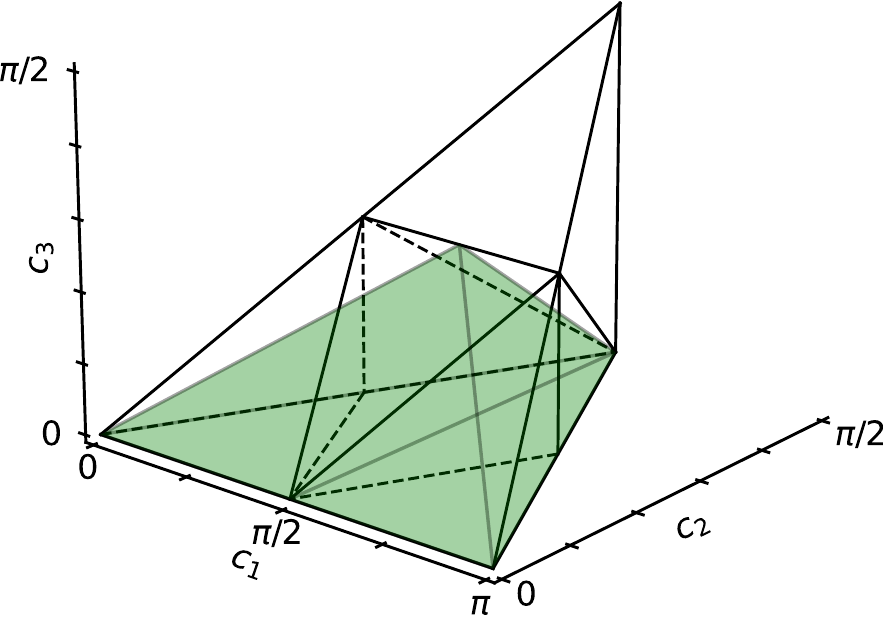}
    \caption{Polytope for \sqiswap{}}  
    \label{fig:poly-sqiswap}
    \end{subfigure}  
    \begin{subfigure}[b]{.45\columnwidth}
    \includegraphics[width=\linewidth]{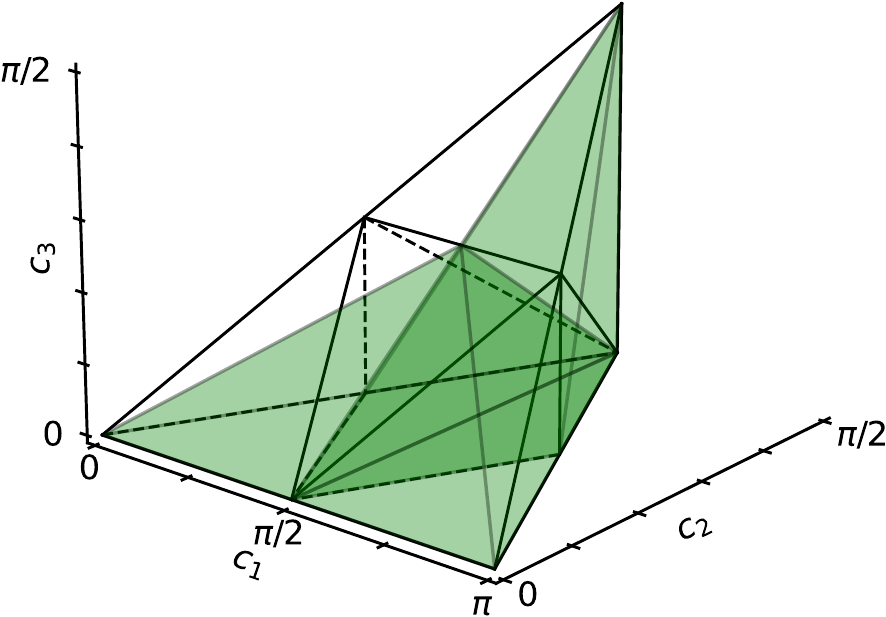}
    \caption{\sqiswap{} with mirror gates}
    \label{fig:poly-sqiswap-mirror}
    \end{subfigure}        
    \caption{Coverage comparison between standard and mirror-inclusive monodromy polytopes for the \cnot{} and \sqiswap{} circuit ansatz with $k=2$. %.  polytope 
    Coverage under the standard scenario is juxtaposed with the coverage when mirror gates are allowed for mirror-inclusive polytopes.}
   % The mirror gate permission visibly increases the coverage, especially for the \sqiswap{} gate.}
    \label{fig:sqiswap_compare_mirrors}
\end{figure}

% Weyl chamber
To reason about decomposition, the requirements of a unitary from the quantum algorithm and the expressible regions within a given basis, a visual representation is \textit{the Weyl chamber}.  The Weyl chamber is a geometric framework derived from Lie Algebra that assigns a unique coordinate to each 2Q unitary operation, invariant under the application of 1Q gates~\cite{zhang2003geometric}.
As 2Q gates represent a unique point in the Weyl chamber, it is possible to describe regions of the Weyl chamber reachable through application of multiple 2Q gates using \textit{monodromy polytopes}~\cite{monodromy}. In particular, these polytopes define the accessible regions of a circuit ansatz for a given basis gate set. These regions, represented as convex polytopes, encapsulate the space of achievable 2Q unitaries within a fixed circuit depth.  Examples of the Weyl chamber shown in Fig.~\ref{fig:sqiswap_compare_mirrors} with regions reachable by a circuit ansatz with $k=2$ \cnot{} and \sqiswap{} gates are shown in Figs.~\ref{fig:poly-cnot} and \ref{fig:poly-sqiswap}, respectively.  

\subsection{Mirror Gates}
\label{subsec:mirror-gates}
% Mirror gate weyl chamber coordinates
%\evm{Revisit - I think there is a typo in the equation. If cannot figure out in time, rewrite equation to be the known equation in canonical form.}
As discussed in Section~\ref{sec:introduction}, a \textit{mirror gate} refers to the resulting gate formed by $U$ composed with a \sw{}.  The \cns{} gate from Fig.~\ref{fig:gate_decompositions} is a special case of mirror gate that has received considerable attention.  It naturally appears in many common circuits For instance, the Toffoli and Fredkin 3Q gates can be decomposed using \cnot{} and \cns{} gates, allowing optimization at the gate-decomposition level~\cite{schuch2003natural,liu2023qcontext, cruz2023shallow}. The \cns{} gate also occurs naturally in stabilizer measurements of error-correcting codes~\cite{schuch2003natural, antipov2023realizing, simakov2022scalable}, in entanglement purification protocols~\cite{tanamoto2008efficient}, and QAOA circuits~\cite{ji2023optimizing, tan2021optimal, hashim2021optimized}. In QFT circuits, the fractional controlled-phase gates can be replaced by \cns{} gates, allowing the QFT circuit to be implemented solely using \iswap{} gates~\cite{wang2011efficient}.  Moreover, the Diamond Gate, a native 4Q gate, can be recast into \cns{} gates, which can be used to build controlled-phase operations~\cite{bahnsen2022application}. 

The transformation of an arbitrary $U$ into its mirror $U'$ has been described in the positive canonical basis, a conventional representation for the Weyl Chamber  coordinates, in Eq.~\ref{eq:mirror-coordinates}~\cite{cross2019validating, peterson2022optimal}. % what do the coverage sets look like?
The two forms deal with the mirrored nature of the Weyl chamber at the midpoint in the line between \cnot{} and \iswap{}.  
\begin{equation}
    (a', b', c') = 
    \begin{cases}
        (\frac{\pi}{4}+c, \frac{\pi}{4}-b, \frac{\pi}{4}-a) & \text{if } a \leq \frac{\pi}{4} \\
        (\frac{\pi}{4}-c, \frac{\pi}{4}-b, a-\frac{\pi}{4}) & \text{else}
    \end{cases}
    \label{eq:mirror-coordinates}
\end{equation}

In this work, we explore the use of monodromy polytopes that are extended to include mirror gates. 
%
% To align with \texttt{monodromy} coordinate conventions we adapt the transformation equations as follows:
%
%
Since polytopes denote the accessible regions within a fixed circuit depth, then if mirroring is permitted, the polytope should also encompass the mirror gates corresponding to every gate within the original region. Thus, we seek to identify the set of unitaries that are accessible, considering an allowance for permutations of output wires, e.g., unitaries that permit a \textit{mirage} \sw{} gate. This is particularly useful in scenarios where the order of the output wires is inconsequential, such as in richly connected topologies found in recently proposed superconducting qubit architectures with local All-to-All (A2A) connectivity~\cite{mckinney2023co}.

To construct a mirror-permitted polytope, we assign a cost of zero to a \sw{} operation, which serves to permute the order of output qubits. Then it is possible to create two polytopes that represent the portion of the Weyl Chamber reachable using a circuit ansatz with a particular value of $k$. For the \cnot{} and \sqiswap{} gates, we depict the $k=2$ case in Figs.~\ref{fig:poly-cnot-mirror} and \ref{fig:poly-sqiswap-mirror}\footnote{The $k=1$ case has 0\% volume as these are only points in the chamber. The $k=3$ case is trivial for both basis gates as they cover 100\% of the Weyl Chamber volume.}. In both the standard polytope and the mirror-permitted polytope for \cnot{}, the planar slices contribute to 0\% volume coverage. In contrast, the \sqiswap{} gate in its standard form covers 79.0\% of the Haar-weighted volume and increases to 94.4\% when mirror gates are utilized. 

\begin{figure}[tbp]
    \centering
    \begin{subfigure}[b]{.45\columnwidth}
       \includegraphics[width=\linewidth]{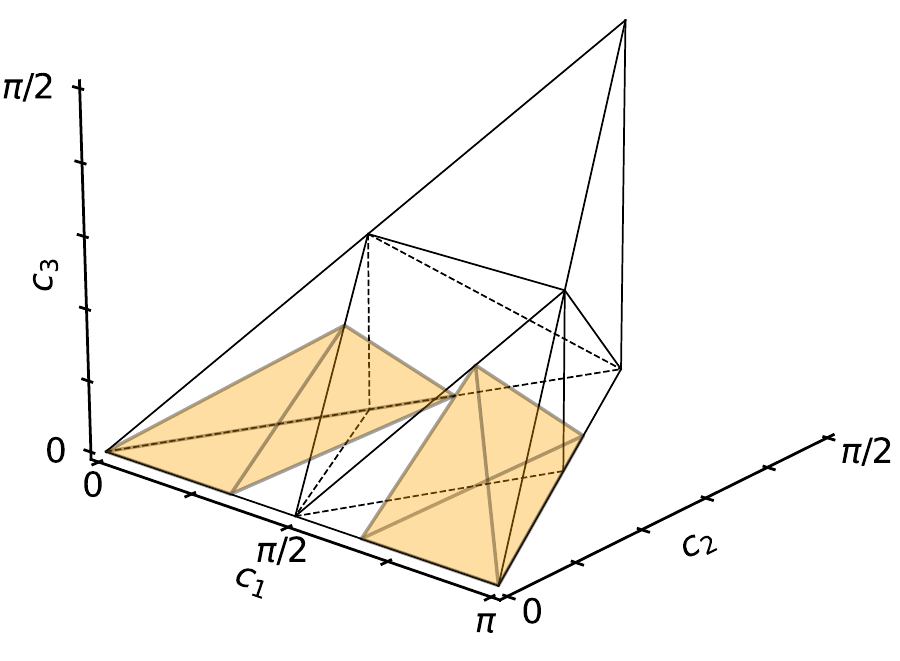}
        \caption{\niswap{3}}
        \label{fig:3rdroot-iswap-polytope}
    \end{subfigure}  
    \begin{subfigure}[b]{.45\columnwidth}
       \includegraphics[width=\linewidth]{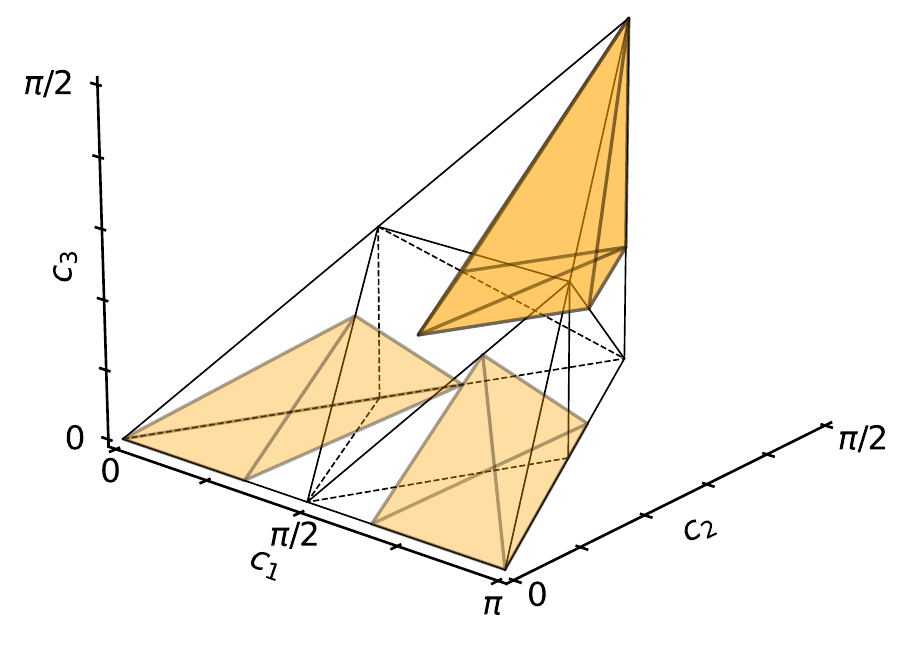}
        \caption{\niswap{3} with mirror gates}
        \label{fig:3rdroot-iswap-polytope-mirror}
    \end{subfigure}        
    \begin{subfigure}[b]{.45\columnwidth}
       \includegraphics[width=\linewidth]{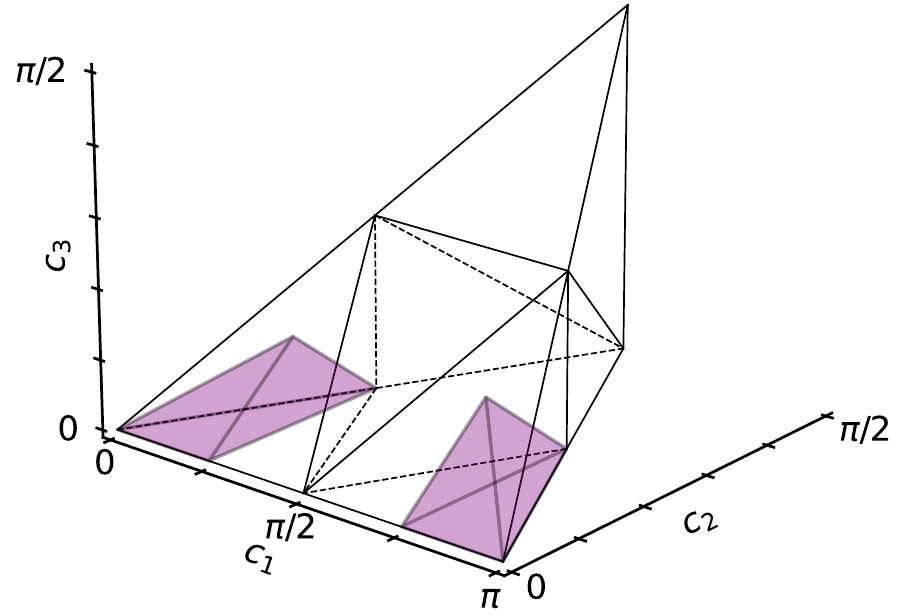}
        \caption{\niswap{4}}
        \label{fig:4throot-iswap-polytope}
    \end{subfigure}  
    \begin{subfigure}[b]{.45\columnwidth}
       \includegraphics[width=\linewidth]{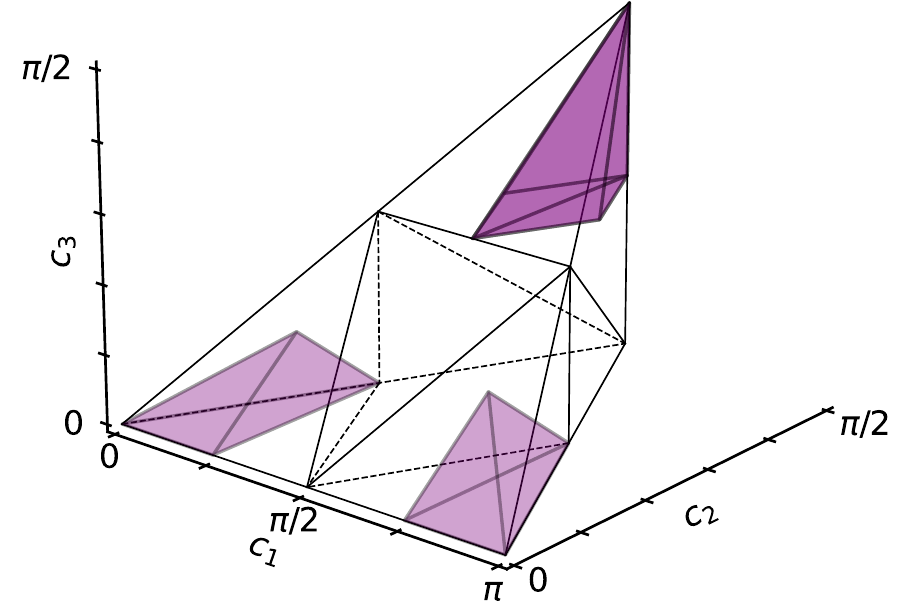}
        \caption{\niswap{4} with mirror gates}
        \label{fig:4throot-iswap-polytope-mirror}
    \end{subfigure}        
    \caption{Coverage comparison between standard and mirror-inclusive monodromy polytopes for the \niswap{3} and \niswap{4} circuit ansatz with $k=2$.}
    \label{fig:reduced-pulse-polytopes-iswap}
\end{figure}

The mirror polytopes from Fig.~\ref{fig:sqiswap_compare_mirrors} intersect. The \cnot{} intersection is a line from \cnot{} to \iswap{},
% , which follows from the \cns{} relationship.  % not true
whereas there is an appreciable region of overlap for \sqiswap{}.  This is another way to illustrate the increased computational power of the \sqiswap{} basis, but it also suggests that perhaps smaller pulse lengths of the \iswap{} family of gates could be particularly useful as we take advantage of mirror gates as shown in Fig.~\ref{fig:reduced-pulse-polytopes-iswap}.  For instance, with \niswap{3} a substantial portion of the Weyl chamber is covered at $k=2$ (Fig.~\ref{fig:3rdroot-iswap-polytope-mirror}).  Even \niswap{4} has useful coverage (Fig.~\ref{fig:4throot-iswap-polytope-mirror}) with the advantage that calibrating such a gate can easily form a \sqiswap{} with two back-to-back pulses.  

Interestingly, while \cnot{} is not reached, many of the \cphase{} gates are in both cases, which is potentially useful in many algorithms.  Furthermore, the maximum cost polytopes  decrease with mirror gates for these partial \iswap{}s; for example, using \niswap{4} traditionally requires up to $k=6$ depth (equivalent to $k=3$ for \sqiswap{}), but with mirroring, the depth never exceeds $k=4$, which guarantees full Weyl Chamber coverage at the equivalent of $k=2$ \sqiswap{}s.  In the next section we quantify this advantage and impact on decomposition fidelity.

\subsection{Computing Fidelity}
% Optimizing Haar score and approximate decomposition Haar score
% Methods for computing Haar scores
One mechanism to quantify the Weyl chamber coverage of a basis gate set is its Haar score, or Haar-average expected circuit cost. This computes the weighted average decomposition cost for a uniformly distributed random 2Q unitary~\cite{zyczkowski1994random}. This cost can be precisely computed using monodromy polytopes. While previous work has shown that mirror gates can enhance Haar scores, this has been done particularly in the context of super-controlled basis gates~\cite{cross2019validating, peterson2022optimal}. Moreover, Qiskit's transpiler implementations have focused on the XX basis to simplify the computation%This methodology has been incorporated into Qiskit, although its application there is specifically tailored for the XX basis
~\cite{qiskit}. 

Generalizing this concept to a broader set of operators requires a function that maps a point outside the polytope coverage region to the nearest point within that region. We use a numerical decomposition method, to optimize an ansatz to the nearest point outside the coverage region to generalize this mirror gates methodology to arbitrary basis gates. Using this approach, we can examine the effects of mirror gates on the Haar scores for \niswap{n} gates.

% Error Model
Practical quantum computing can take advantage of approximate decompositions that may lead to higher overall fidelity because they require fewer applications of the noisy basis gate~\cite{lao2021designing, mckinney2023co, rakyta2022efficient}. In other words the infidelity from the approximation is less than the infidelity from the increased noise resulting from the additional complexity of circuit necessary to compute the exact result.  To measure this, we use an error model proposed in previous work~\cite{gokhale2021faster, mckinney2023parallel} that identifies decoherence over time as the main source of infidelity. This model defines the fidelity of a gate, $F_Q$, in terms of an exponential decay relative to the gate duration and the qubit's T1 relaxation rate as described in Eq.~\ref{eq:decoherence}~\cite{koch2007charge}.

\begin{equation}
F_Q = e^{-\text{Gate Duration}/\text{Qubit Lifetime}}
\label{eq:decoherence}
\end{equation}

Using this model, the decomposition task becomes an optimization problem that balances circuit fidelity and decomposition fidelity. The total fidelity, a product of these two fidelities, governs the acceptance threshold for a given circuit. This decomposition error tolerance threshold can be intuitively understood as expanding the volume of each of the coverage sets~\cite{javadi2023improving}.

% Monte Carlo method for approximate decomposition Haar score
To compute updated Haar scores that factor in both mirror gates and approximate decompositions, we use Monte Carlo sampling of unitary targets from the Haar distribution, and verify decomposition fidelity using numerical decomposition. First, we calculate circuit infidelity from the exact decomposition solution before subsequently checking if any cheaper polytopes (corresponding to higher fidelity circuits) can approximate the target circuit within the prescribed total fidelity threshold. The overall process is encapsulated in Algorithm~\ref{alg:haar-mc}.

% Monte Carlo Convergence Figure
\begin{algorithm}[tbp]
\caption{Monte Carlo for approximate decomposition Haar scores}
\label{alg:haar-mc}
\begin{algorithmic}[1]
\footnotesize
\Procedure{ApproxGateCosts}{$N$}
    \State $TotalCost \gets 0$
    \For{$i \text{ in range}(N)$}
        \State $Target \gets \text{HaarSample()}$
        \For{each $P \text{ in } Set$}
            \If{$P \text{ contains } Target$}
                \State Compute $ExactCost$ and $FidThreshold$
            \EndIf
        \EndFor
        \State $BestCost \gets ExactCost$
        \For{each $P \text{ in } CheaperSet$}
            \State $Cost \gets \text{Optimize}(P, Target, FidThreshold)$
            \If{$Cost \neq \text{None}$}
                \State $BestCost \gets \min(BestCost, Cost)$
            \EndIf
        \EndFor
        \State $TotalCost \gets TotalCost + BestCost$
    \EndFor
    \State \Return $TotalCost / N$
\EndProcedure
\end{algorithmic}
\end{algorithm}

% Monte Carlo convergences.
We perform our calculations based on the assumption that the gate \iswap{}, which carries a normalized unit cost of 1.0 with fidelity of 99\%~\cite{zhou2023realizing}. Consequently, fractional \niswap{n} gates, which have shorter unit time costs (\textit{i.e.}, less decoherence time, e.g., \sqiswap{} with 0.5) have proportionately adjusted fidelities. The fidelity of the final decomposition obtained through approximate decomposition is recorded for each iteration. 

\begin{figure}[tbp]
    \centering
    \includegraphics[width=\columnwidth]{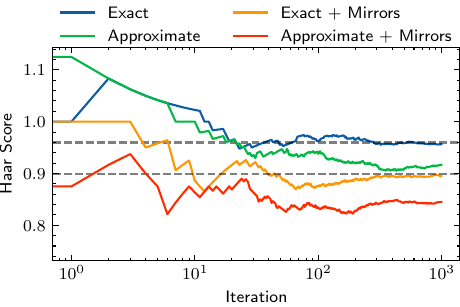}
    \caption{The convergence of the Haar Score for \niswap{4} across 1000 iterations, employing different optimization strategies. Horizontal lines denote the exact values derived through polytope integration.}
    \label{fig:montecarlo_convergence}
\end{figure}
The Monte Carlo convergence for \niswap{4}, subject to 1000 iterations, is illustrated in Fig.~\ref{fig:montecarlo_convergence}. It can be noted that the Exact and Exact + Mirrors solutions converge successfully at the theoretically computed values illustrated with dotted lines in the figure.  This gives confidence that the decomposition that allows for approximate solutions also reasonably converges. Noting, a lower Haar score is more desirable, the approximate solution without mirrors nearly reaches the exact solution that contains mirror gates.  By combining both, approximation and mirror gates, the Haar score improves from 0.9 to under 0.85.

% Haar score and Fidelity Table Results
\begin{table}[tbp]
  \centering
  \resizebox{\columnwidth}{!}{%
    \begin{tabular}{|c|c|c|c|c|}
      \hline
      Basis Gate & Haar & Fidelity & Mirror Haar & Mirror Fidelity \\ \hline
      \niswap{2} & 1.105 & 0.9890 & 1.029 & 0.9897 \\ \hline
      \niswap{3} & 0.9907 & 0.9901 & 0.9545 & 0.9904 \\ \hline
      \niswap{4} & 0.9599 & 0.9904 & 0.8997 & 0.9910 \\ \hline
    \end{tabular}
  }
\caption{Comparison of Haar scores and corresponding average fidelities for different basis gates, both with and without mirror decomposition.}
\label{tbl:haar_mirror}

  \centering
  \resizebox{\columnwidth}{!}{%
    \begin{tabular}{|c|c|c|c|c|}
      \hline
      Basis Gate & Haar & Fidelity & Mirror Haar & Mirror Fidelity\\ \hline
      \niswap{2} & 1.031 & 0.9895 & 0.9950 & 0.9899 \\ \hline  
      \niswap{3} & 0.9433 & 0.9904 & 0.8900 & 0.9908 \\ \hline
      \niswap{4} & 0.9165 & 0.9906 & 0.8453 & 0.9913 \\ \hline
    \end{tabular}
  }
\caption{Comparison of Haar scores and corresponding average fidelities for different basis gates, taking into account approximate decompositions.}
\label{tbl:approx_mirror}
\end{table}
The average total fidelities and Haar score for exact decomposition, both with and without allowed mirror gates, are recorded in Table~\ref{tbl:haar_mirror}. Similarly, the results that allow for approximate decomposition when it improves fidelity and Haar score, again with and without allowed mirror gates, are presented in Table~\ref{tbl:approx_mirror}.

% Results
The observed fidelity improvements are particularly noteworthy, given that they stem solely from compiler-level optimizations, without necessitating experimental adjustments (i.e., a hardware change). For example, \textbf{using \sqiswap{} as the basis gate, the transition to approximate mirror decomposition provides an 8.8\% relative decrease in total infidelity, and a 9.4\% decrease when \niswap{4} is used}. While smaller fractional basis gates can bolster Haar scores, previous work reveals that as these basis fractions diminish, the contribution of interleaving single-qubit gates becomes increasingly significant. Furthermore, the primary target of interest, \cnot{}, will not experience further improvements with progressively smaller fractional gates \iswap{}, indicating a limit to this strategy~\cite{mckinney2023parallel}. Hence, this further supports \sqiswap{} as a basis gate of choice. 

\subsection{Mirroring on Restricted Topologies}

% Discussion about the limitations of using mirror gates in restricted topologies and the need for a routing algorithm
Although mirror gates will reduce gate decomposition costs in A2A topologies, in practical topologies with less rich connectivity, they may inadvertently add unfavorable qubit permutations, which would need to be undone using a \sw{}, nullifying any benefit from the cheaper mirror decomposition. A transpilation algorithm is essential to find optimal use of mirror gates, taking into account subsequent gates to minimize both gate decomposition costs and \sw{} operations.

\begin{figure}
    \centering
    \includegraphics[width=\columnwidth]{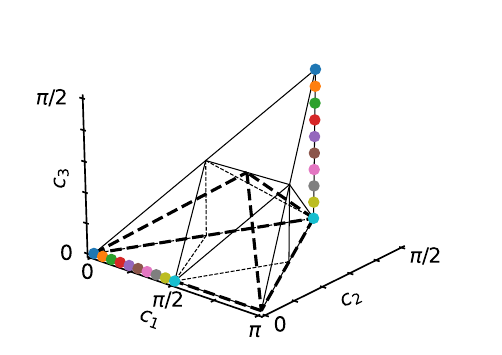}
    \caption{\cphase{} gates and their respective mirrors into \pswap{} plotted against \sqiswap{} $k=2$ coverage.}
    \label{fig:cphase-mirrors}
\end{figure}

% investigate special case of CPhase gates.
Quantum computing algorithms frequently involve controlled-phase or \cphase{} gates. Using Eq.~\ref{eq:mirror-coordinates}, it is possible to mirror the \cphase{} family into the parametric-\sw{} family~\cite{crooks2020gates}, depicted in Fig.~\ref{fig:cphase-mirrors} in the Weyl chamber. The figure also superimposes the $k=2$ coverage region for \sqiswap{} in bold dotted lines.  This region contains the \cphase{} gates, but not the \pswap{} gates. However, the $k=2$ coverage region contains both \cnot{} and its mirror (recall Fig.~\ref{fig:gate_decompositions}, \cns{} or \iswap{}).  Thus, decomposition cost would remain unchanged upon mirroring. Although \pswap{} has a higher decomposition cost ($k=3$), it would be preferred if it could eliminate a \sw{} operation. For these cases, the emphasis is absorbing \sw{} operations with a secondary objective to accomplish this without any decomposition cost penalty. Another perspective is that mirror gates can provide a \sw{} when it is helpful for improved routing, particularly when it does not increase decomposition cost.

% transition into routing algorithm section
In the next section, we will detail our routing algorithm, MIRAGE, which strategically uses mirror gates based on the broader context of the quantum circuit. The goal is not only to minimize costs by mirroring every possible gate, but to strategically place and select the decomposition of the mirrors considering downstream operations, topology constraints, and the potential to absorb \sw{} operations.

\section{MIRAGE}
In this section, we introduce MIRAGE, a collaboratively designed transpilation algorithm that collectively examines \sw{} insertion and decomposition.  MIRAGE can exploit the intrinsic efficiency of \sqiswap{} gate, which decomposes both \cnot{} and \cns{} at the same cost, an advantage that is not available to the traditional \cnot{} basis (Fig.~\ref{fig:gate_decompositions}).  However, conceptually, MIRAGE can be used for any basis gate.  Using monodromy polytopes and mirror and approximation based decomposition discussed in Section~\ref{sec:mirror-gate-decomposition}, MIRAGE can select between gates and their mirror form to determine the appropriate choice for implementation in the circuit.

MIRAGE is a heuristic approach with greedy characteristics similar to the prior state-of-the-art transpilation passes, such as SABRE~\cite{li2019tackling}.  However, unlike previous approaches, MIRAGE considers both routing and decomposition cost when determining both \sw{} placement and use of mirror gates.  However, because MIRAGE fundamentally maintains a greedy approach, we consider several strategies that allow MIRAGE to avoid getting trapped into poorly performing situations, such as finding a local minimum far from the global optimum.

First, when considering a gate or swap choice, we evaluate the potential impact on circuit depth rather than solely \sw{} depth as is considered in prior work.  Second, we propose mirror gate aggression levels which use different circuit depth thresholds to determine whether to insert a mirror gate.  %Third, we explore the use of simulated annealing to make the algorithm less dependent on the initial conditions of the circuit placement and the choice of aggression level to find a near optimal solution.
MIRAGE is implemented using the SABRE workflow included in Qiskit with several key modifications.  We describe these next.

% FIGURE MIRAGE workflow
\begin{figure*}
\centering
\includegraphics[width=\textwidth]{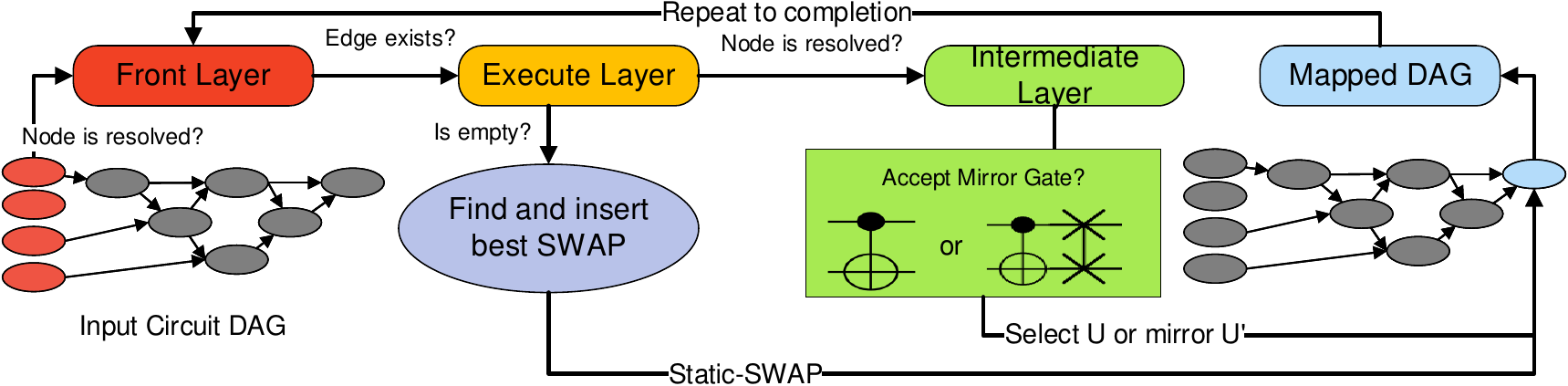}
\caption{MIRAGE Workflow}
\label{fig:Modified_SABRE_Algorithm}
\end{figure*}
\subsection{MIRAGE Workflow Details}

% orginal SABRE workflow
The SABRE workflow provides several key data structures and a optimization flow that we inherit into MIRAGE. The circuit is represented as a directed acyclic graph (DAG).  The gate nodes from the DAG are organized into a front layer, an execution layer, and a mapped layer. The mapped layer holds the portion of the DAG that has be mapped to the hardware.  The front layer holds nodes with resolved dependencies, while the execution layer accommodates nodes that can be executed based on the current physical qubit layout. The fully processed nodes are transferred to the mapped layer into the mapped DAG. 

Gates move sequentially from the unmapped DAG to the front layer to the mapped layer, a process guided by the resolution of their predecessors. In other words, a gate can only progress to a stage once all its predecessor nodes have advanced past that stage. When the execution layer is empty, but unresolved gates remain in the front layer, SABRE selects some number of \sw{} gates that are inserted into the execute layer with a goal of minimizing the heuristic cost function, designed to minimize topological distance between upcoming qubits and promote parallelism. This parallelism comes from selecting \sw{} gates that can be executed at the same time to concurrently move data towards both inputs of the gate being mapped.  Once sufficient \sw{} gates are introduced, this stalled node in the front layer \textit{advances}. After all gates in the front layer advance a complete mapping of all gates is accomplished in the mapped DAG.

In MIRAGE we retain the front, execute, and mapped layers. 
However, we add an intermediate layer between the execute and mapped layer as shown in Fig.~\ref{fig:Modified_SABRE_Algorithm}.  The intermediate layer is used to handle every 2Q gate from the algorithm  leaving the execution layer. The transpiler pass evaluates whether to accept the mirror gate in its place, then ultimately sends the chosen operation to the mapped layer. 
Note: in the original SABRE workflow, these 2Q gates are not decomposed to the basis gate of the target machine.  However, in MIRAGE, while the gate is not decomposed, a decomposition estimate is %created using monodromy polytopes.   Thus, when a mirror change is 
considered in the intermediate layer and/or execution layer to improve the estimation of mirror or \sw{} choices over pure topological distance and gate depth. %, MIRAGE considers whether this gate can be combined with other gates for decomposition rather than just counting gate depth.

\begin{figure}
    \centering
    \begin{subfigure}[b]{\columnwidth}
        \centering
    \resizebox{.7\columnwidth}{!} {
    \begin{quantikz}
    q_0 & \ctrl{1} & \ctrl{2} & & \ctrl{3} & & &
    \\
    q_1 & \targ{} & & \ctrl{1} & & \ctrl{2} & & 
    \\
    q_2 & & \targ{} & \targ{} & & & \ctrl{1} &
    \\
    q_3 & & & & \targ{} & \targ{} & \targ{} &
    \end{quantikz}        
    }
        \caption{TwoLocal (full), 4 qubits}
        \label{fig:twolocal-base}
    \end{subfigure}
        
    \begin{subfigure}[b]{\columnwidth}
        \centering
    \resizebox{.8\columnwidth}{!} {
    \begin{quantikz}
    q_0 \rightarrow 0 & \slice{2} & \targ{} \slice{4} & \slice{7} & \targ{} & \slice{9} & \slice{11} & \swap{1} \slice{14} & \slice{16} &
    \\
    q_1 \rightarrow 1 & \ctrl{1} & \ctrl{-1} & \swap{1} & \ctrl{-1} & & \ctrl{1} & \targX{} & \ctrl{1} &
    \\
    q_2 \rightarrow 2 & \targ{} & & \targX{} & \ctrl{1} & \swap{1} & \targ{} & & \targ{} &
    \\
    q_3 \rightarrow 3 & & & & \targ{} & \targX{} & & & &
    \end{quantikz}
    }

        \caption{TwoLocal (full), 4-qubit line topology, Qiskit level 3}
        \label{fig:twolocal-qiskit}
    \end{subfigure}

    \begin{subfigure}[b]{\columnwidth}
        \centering
    \resizebox{\columnwidth}{!} {
    \begin{quantikz}
    q_0 & \targ{} \slice{2} & \gate{U \atop {0,0,\frac{-\pi}{2}}}\!\!\! 
    & \slice{4} & & \iSwap{1} \slice{6} & \gate{U \atop {\frac{\pi}{2},0,\pi}}\!\!\! & \slice{8} & & \ctrl{1} \slice{10} &
    \\
    q_1 & \ctrl{-1} & \gate{U \atop {0,0,\frac{-\pi}{2}}}\!\!\! & \iSwap{1} & \gate{U \atop {0,0,\frac{-\pi}{2}}}\!\!\! & \targiS{} & 
    \gate{U \atop {0,0,\frac{-\pi}{2}}}\!\!\! & \iSwap{1} &
    \gate{U \atop {\frac{\pi}{2},0,\pi}}\!\!\! & \targ{} &
    \\
    q_2 & \gate{U \atop {\frac{\pi}{2},\frac{-\pi}{2},\pi}}\!\!\!
    & & \targiS{} & \gate{U \atop {0,0,\frac{-\pi}{2}}}\!\!\! & \iSwap{1} & 
    \gate{U \atop {0,0,\frac{-\pi}{2}}}\!\!\! & \targiS{} & & &
    \\
    q_3 & \gate{U \atop {\frac{\pi}{2},\frac{-\pi}{2},\pi}}\!\!\!
    & & & & \targiS{} & &
    & & &
    \end{quantikz}
    }
        \caption{TwoLocal (full), 4-qubit line topology, MIRAGE}
        \label{fig:twolocal-vswap}
    \end{subfigure}
    
    \caption{Comparison of TwoLocal circuits}
    \label{fig:twolocal}
\end{figure}
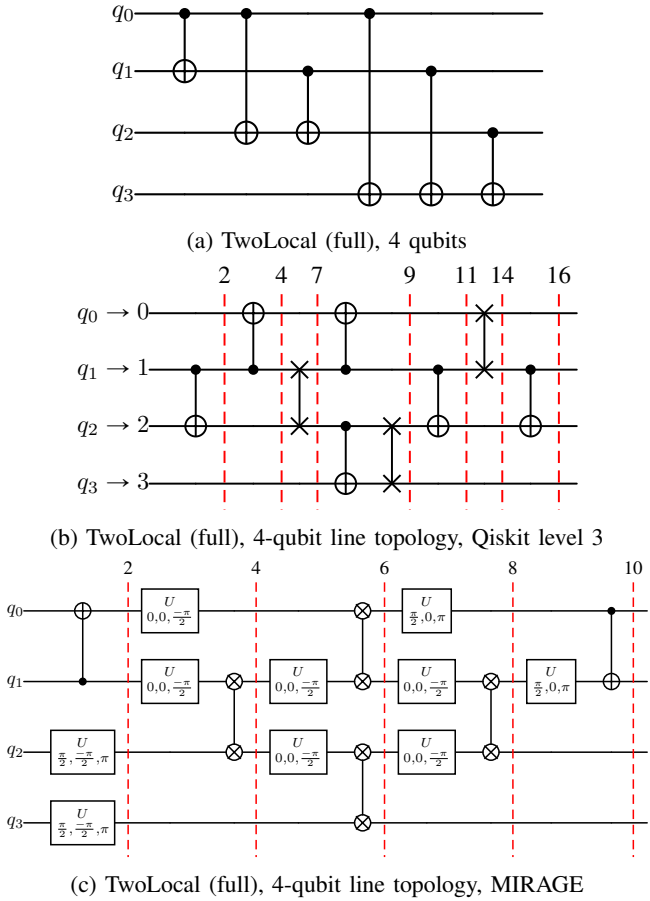

The impact of MIRAGE is illustrated in Fig.~\ref{fig:twolocal} targeting a \sqiswap{} machine considering a fully entangling ansatz--a common structure in quantum machine learning and optimization algorithms (Fig.~\ref{fig:twolocal-base}). When this circuit is mapped to a line topology and optimized using Qiskit at level 3, the circuit depth amounts to 16 \sqiswap{} basis gates using 3 \sw{} gates (Fig.~\ref{fig:twolocal-qiskit}). To interpret this pulse count, with \sqiswap{} basis gates it requires $k=2$ to implement a \cnot{} and $k=3$ to implement a \sw{}.  Note, between pulses 7 and 9, the \cns{} gate between $q_2$ and $q_3$ can be realized as an \iswap{} which is equivalent to \sqiswap{} with $k=2$.  

In contrast, using MIRAGE, the \sw{}s are better absorbed into gates.  By inverting the first \cnot{} the next \cnot is between $q_1$ and $q_2$ eliminating a \sw{}.  Replacing that \cnot{} with a mirror sets up doing the third and fourth \cnot{} in parallel because both gates are independent and both are one step away.  Ultimately, MIRAGE finds the same circuit functionality with only 10 \sqiswap{} gates and no \sw{} gates, underscoring the potential optimization (Fig.~\ref{fig:twolocal-vswap}).

% \akj{I think this is now redundant}
% Like the original SABRE workflow, in MIRAGE, a DAG node moves to the front layer only after all its predecessors have exited (been mapped to the system). This rule also applies to transitions into the intermediate layer. Importantly, when a node resides in the intermediate layer, its predecessors must remain in the front layer to ensure accurate calculation of the heuristic cost based on the current layout's topological distance over nodes in the front and lookahead layers. When there's a halt in the processing of nodes in the front layer, we proceed to process all nodes in the intermediate layer. This involves examining the gate nodes to determine whether replacing the unitary $U$ with its mirror $U'$ and swapping qubit output reduces the upcoming topological distance. If so, we push $U'$ to the mapped DAG and update the layout as if a \sw{} operation had occurred.  If both layers are stalled or empty, we invoke the original SABRE to provide the next \sw{} gate. 

% \subsection{MIRAGE}
% % Introduction of MIRAGE
% \evm{Incomplete sentence?}
% an enhancement of the SABRE algorithm that exploits the unique properties of the \sqiswap{} basis gate. Our approach incorporates an  DAG (see Fig.~\ref{fig:Modified_SABRE_Algorithm}). This approach leverages the intrinsic efficiency of \sqiswap{}, which decomposes both \cnot{} and \cns{} at the same cost, an advantage that is not available to the traditional \cnot{} basis (Fig.~\ref{fig:gate_decompositions}).

%The MIRAGE Workflow

% Importance of novel contributions

\subsection{Post-Selection Metric}
\label{sec:post-select-metric}
% SABRE post-selects with SWAP count
An important advancement of MIRAGE is the selection metric to choose the optimal routing among multiple independent trials. The original SABRE implementation relies on a decay factor to promote parallelism, discouraging the use of \sw{} gates on qubits that have recently undergone a \sw{}.  However, when performing multiple routing trials, the tracked metric is the total number of induced gates \sw{}, which is less closely related to the depth of the circuit~\cite{li2019tackling}. This disconnect arises because the transpiler cannot reason about depth until it performs decomposition into the basis gate.

% MIRAGE post-selects with depth
MIRAGE, addresses this by enabling the estimation of circuit depth without the need for actual decomposition. % Calculating Depth Metric using Monodromy
This is accomplished using monodromy polytopes to rapidly assess circuit costs. The minimum-cost circuit polytope that contains the unitary target is identified as discussed in Section~\ref{sec:mirror-gate-decomposition}. We iterate the coverage set in until we find a coverage region containing the edge's 2Q gate. Thus, the depth metric is calculated using %\texttt{retworkx}'s 
the longest DAG path with a custom weight function assigned to decomposition cost.  Total gate counts are calculated similarly, summed over all nodes.

% Basis gate agnostic
While exact methods for decomposing into \cnot{} and \sqiswap{} exist, MIRAGE is designed to operate independently of any specific decomposition strategy. The actual decomposition can be specified later using either exact or numerical methods. 
While we continue to use decay when choosing individual \sw{}s, the post-selection process is about choosing between routes across the independent trials. 

\subsection{Overcoming Local Minimas}
\begin{figure}[tbp]
    \centering
    \includegraphics[width=\linewidth]{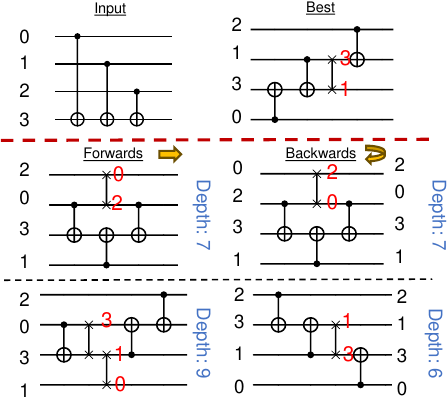}
    \caption{Two routing trials from the same initial layout demonstrate the challenges of local decision-making. The top route, despite an initial optimal choice, ends in a local minima, while the bottom, through an initial sub-optimal choice, finds the best solution.}
    \label{fig:bv-edge-case}
\end{figure}

Unfortunately, even with the best possible lookaheads, greedy algorithms have a propensity to get stuck in local minima that can be quite a bit worse than the optimal solution. 
We address the challenge of local minima that can obstruct the optimization process by noting that the algorithm's sensitivity to initial placement and cost decision making can lead to certain cases where the initial layout enters a non-converging cycle, as shown in Fig.~\ref{fig:bv-edge-case}.

% FIGURE BV Circuit Edge Case

In this example, which is a subset of the circuit from Fig.~\ref{fig:twolocal-base}, the qubits are reordered, so that the first gate requires no \sw{}s.  One choice is to place a \sw{} between the top qubits while executing a \cnot{} between the bottom qubits, followed by a \cnot{}.  This approach has a depth of seven pulses.  However, the minimal has to allow two swap gates be added to further reorder the qubits.  While one \sw{} can be combined into a \cns{} gate, the total pulses becomes nine, except, that this leads to the solution on the bottom right, by reversing the \sw{} gates, the \cns{} gate moves to the middle gate, creating a solution with only six pulses, while the local minima got stuck at seven.

% strategy is compare between aggressions and simualted annealing
% We evaluated two strategies to mitigate this issue: variable aggression settings and simulated annealing.
To address this, we use a technique described as \textit{mirror aggression settings} to dictate the likelihood of accepting a mirror gate decomposition in the intermediate layer. Specifically, we define four aggression levels: level 0, a mirror gate is never accepted; level 1, a mirror gate is accepted if it lowers the cost; level 2, a mirror gate is accepted if it either lowers or maintains the cost; and level 3, a mirror gate is always accepted. %In contrast, simulated annealing offers a comprehensive alternative. This probabilistic technique, used to approximate the global optimum of a given function, gradually decreases the acceptance rate of sub-optimal solutions over successive iterations.

\begin{figure}[tbp]
    \centering
    \includegraphics[width=\columnwidth]{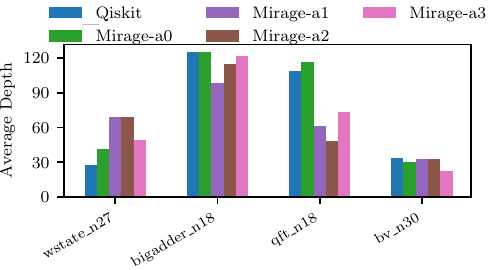}
    \caption{Results of independent configuration trials with different aggression settings.}
    \label{fig:aggression-levels}
\end{figure}

In practice, different circuits perform well with different aggression levels.  To evaluate the impact of fixed aggression settings, we conducted tests using each level of aggression (Figure~\ref{fig:aggression-levels}). We selected a subset of circuits to demonstrate that \textbf{no single strategy is universally optimal}. The results support the use of a mix of aggression settings, allowing MIRAGE to handle a wide range of circuits and topologies. Based on these representative circuit trials, we distribute the routing trials across aggressions as follows: 5\% at level 0, 45\% at level 1, 45\% at level 2, and 5\% at level 3. This approach, with some small effort on the edge cases (levels 0 and 3), with the majority of effort on the metric-based choices improves the effectiveness of various circuits on different arbitrary initial layouts.

%DATA, results when an aggression fixed at single value.

\begin{algorithm}
\caption{Mirror Gate Acceptance Function}
\footnotesize
\begin{algorithmic}[1]
\State \textbf{Input:} cost\_current, cost\_trial, aggression%, temperature
\State \textbf{Output:} Boolean accept mirror gate

\If {aggression = 0}
    \State \textbf{return} False
\ElsIf {aggression = 1 and cost\_trial \textless cost\_current}
    \State \textbf{return} True
\ElsIf {aggression = 2 and cost\_trial $\leq$ cost\_current}
    \State \textbf{return} True
\ElsIf {aggression = 3}
    \State \textbf{return} True
% \ElsIf {cost\_trial $\geq$ cost\_current}
%     \State Compute $P = exp((cost\_current - cost\_trial) / T)$
%     \If {P \textgreater random\_number}
%         \State \textbf{return} True
%     \EndIf
\EndIf
\State \textbf{return} False
\end{algorithmic}
\label{alg:probability-accept}
\end{algorithm}

% section conclusion
Algorithm~\ref{alg:probability-accept} describes the final MIRAGE algorithm that combines variable aggression settings.
%and simulated annealing to improve the success rate of any given layout initialization. 
This approach optimizes circuit compression and reduces the number of layout trials required. Future work could explore further parameter fine-tuning. 

\section{Experimental Setup}
% FLow: Qiskit/Tool setup -> Benchmarks -> Performance Metrics
To evaluate the performance and effectiveness of MIRAGE we implemented the MIRAGE algorithm in \texttt{qiskit-terra} 0.24.2 and used this flow in our experiments. To compute monodromy polytopes, we used the Qiskit-Extension tool, \texttt{monodromy}~\cite{monodromy} which we adapted to compute cost-weighted integration for both the standard and mirror-permitted coverage polytopes.

Our setup does not exceed the parameters in the default SABRE workflow: a lookahead window size ($|E|$) of 20, a window weight ($W_E$) of 0.5, and a decay rate of 0.001, with a reset after every five steps or gate mapping. We addressed a limitation in Qiskit's SABRELayout, which omits independent layout trials when a custom routing pass, like MIRAGE, is specified. We modified SABRELayout to adhere to the original SABRE configuration when MIRAGE is specified: 20 independent layout trials, each with 4 forward and backward routing passes, each iteration routed independently 20 times. 

% Changes we have to make to Qiskit to get things working
To establish a baseline, we added the \cnot{} and \sw{} decomposition rules to the session equivalence library, as Qiskit lacks support for \sqiswap{}. However, this was only done for final circuit output as our MIRAGE cost functions consolidate all blocks before calculating costs using \texttt{monodromy}.  Moreover, for decomposition to an arbitrary basis, \texttt{monodromy} identifies the minimum cost polytope that contains a target gate in its region. The resulting circuit ansatz formed from the basis gates is then  optimized by fitting the 1Q gate parameters using a numerical optimizer. %This tool is particularly beneficial for exploring instruction sets, as it can accommodate any basis gate or even an over-complete basis set. In our work, we leverage the versatility of \texttt{monodromy} for our analysis of mirror basis gates, demonstrating the robustness and flexibility of this tool.

When executing the transpiler, the pass manager conducts input cleaning, which includes unrolling gates with more than two qubits, removing \sw{}s , barriers, and identity gates. We consolidate all consecutive unitary blocks, ensuring MIRAGE operates solely on 2Q gates. We then check if an implementation with no \sw{} gates can be found using VF2Layout.  We invoke MIRAGE (or SABRE) if no \sw{} gate free placement can be found. Subsequently, we incorporate Qiskit's remaining optimizations and reconsolidate the circuit.

Lastly, for speed comparisons, we compared MIRAGE to the most recent version of SABRE in Python. While a version of SABRE has been ported to Rust, this made it impractical to determine the impact of MIRAGE on runtime. 

% Benchmark Selection
We evaluate MIRAGE's performance using circuits from QASMBench~\cite{li2020qasmbench} and MQTBench~\cite{quetschlich2023mqtbench}. These benchmarks are selected for their relevance to NISQ devices and their need for $>0$ \sw{} gates. This is crucial because our transpiler, like the stock Qiskit implementation, checks using VF2Layout if an optimal mapping exists that requires no \sw{} gates. Consequently, for circuits like GHZ or any linear ansatz VQA, both transpilers would behave identically, and neither SABRE nor MIRAGE would be invoked.

% TABLE Circuit Benchmarks
\begin{table}[htbp]
    \centering
    \resizebox{\columnwidth}{!}{%
    \begin{tabular}{|l|c|c|c|}
        \hline
        \textbf{Name} & \textbf{Qubits} & \textbf{2Q Gates} & \textbf{Class}\\
        \hline
        wstate~\cite{li2020qasmbench} & 27 & 52 &  Entanglement\\
        \hline
        qftentangled~\cite{quetschlich2023mqtbench} & 16 & 279 & Hidden Subgroup\\
        \hline
        qpeexact~\cite{quetschlich2023mqtbench} & 16 & 261 & Hidden Subgroup\\
        \hline
        ae~\cite{quetschlich2023mqtbench} & 16 & 240 & Hidden Subgroup\\
        \hline
        qft~\cite{quetschlich2023mqtbench} & 18 & 306 & Hidden Subgroup \\
        \hline
        bv~\cite{li2020qasmbench} & 30 & 18 & Hidden Subgroup \\
        \hline
        multiplier~\cite{li2020qasmbench} & 15 & 246 &  Arithmetic\\
        \hline
        bigadder~\cite{li2020qasmbench} & 18 & 130 & Arithmetic \\
        \hline
        qec9xz~\cite{li2020qasmbench} & 17 & 32 & EC\\
        \hline
        seca~\cite{li2020qasmbench} & 11 & 84 & EC \\
        \hline
        qram~\cite{li2020qasmbench} & 20 & 92 & Memory \\
        \hline
        sat~\cite{li2020qasmbench} & 11 & 252 & QML\\
        \hline
        portfolioqaoa~\cite{quetschlich2023mqtbench} & 16 & 720 & QML\\
        \hline
        knn~\cite{li2020qasmbench} & 25 & 96 &  QML \\
        \hline
        swap\_test~\cite{li2020qasmbench} & 25 & 96 &  QML\\
        \hline
    \end{tabular}%
    }
    \caption{Selected circuit benchmarks}
    \label{table:hpca_circuits}
\end{table}

% Performance Metrics
We measure transpiler success using circuit costs via normalized critical path durations. In our convention, an \iswap{} gate has a time cost of 1.0, and a \sqiswap{} has 0.5, as defined in Section~\ref{sec:mirror-gate-decomposition}. We report the geometric mean of circuit depth across 5 instances for each experiment. Our focus on reducing circuit depth, compared to Qiskit SABRE, underscores MIRAGE's utility as a straightforward yet significant enhancement to the existing Qiskit routing stage. 

% \section{MIRAGE Configurations}
% % motivation: what is the transpiler pass manager actually doing?
% In this section, we outline the specific configuration choices and modifications made to optimize MIRAGE.

% Overview of the Pass Manager

% FIGURE compare Min-Swaps to Min-Depth post-selection

\section{Results}
\label{sec:results}
In this section we examine the impact of the MIRAGE transpilation approach.  First we explore the impact of circuit depth comparison against number of gates, then we examine the effectiveness of the MIRAGE optimization for different common quantum machine topologies, and finally we consider the runtime impact of MIRAGE against prior work.  

\subsection{Impact of Depth Comparison Metric}
\begin{figure}[tbp]
    \centering
    \includegraphics[width=\columnwidth]{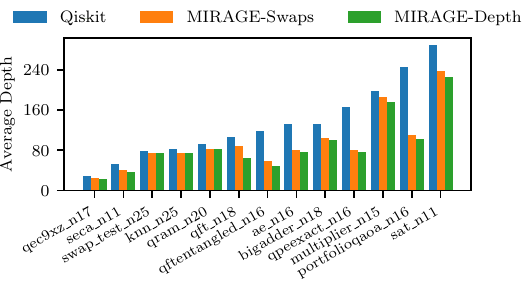}
    \caption{Average circuit depth comparison for Qiskit, Mirage-SWAPs, and Mirage-Depth post-selection strategies.}
    \label{fig:post-select}
\end{figure}

% Results + Analysis
Previous transpilers tracked the circuit depth in terms of number of \sw{} gates added to the circuit.  MIRAGE improves on this by using a circuit depth metric to determine the best result as discussed in Section~\ref{sec:post-select-metric}.  In Figure~\ref{fig:post-select} we compared to the stock Qiskit SABRE, which determines the best result in of the fewest \sw{} gates added, with MIRAGE using the same metric as well as our circuit depth cost function that more efficiently promotes parallelism.  Our results indicate that both methods using MIRAGE find an improvement, due to inclusion of mirror gates. However, when we optimize for minimum swaps, we find an average depth reduction of 24.1\%, and when we optimize for depth, we get \textbf{an additional 7.5\%} benefit, leading to a total average depth reduction of 29.5\%. Interestingly, the total number of gates is mostly unchanged (an increase of 0.4\%), indicating that changing the selection metric is responsible for finding more parallelism. 

% Results Hook
\subsection{MIRAGE for Common Quantum Machine Topologies}
\begin{figure*}[t!]
    \centering
    \begin{subfigure}[b]{\columnwidth}
        \centering
        \includegraphics[width=\columnwidth]{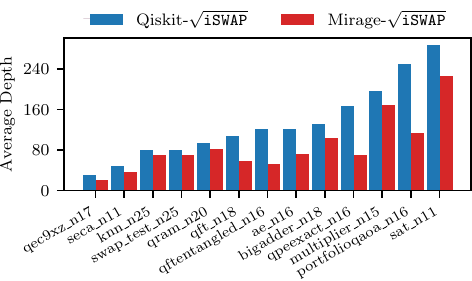}
        \caption{Critical path depth for Heavy-Hex topology}
        \label{fig:hex-depth}
    \end{subfigure}
    \begin{subfigure}[b]{\columnwidth}
        \centering
        \includegraphics[width=\columnwidth]{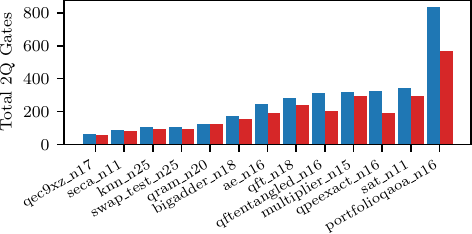}
        \caption{Total gate cost for Heavy-Hex topology}
        \label{fig:hex-total}
    \end{subfigure}

    \begin{subfigure}[b]{\columnwidth}
        \centering
        \includegraphics[width=\columnwidth]{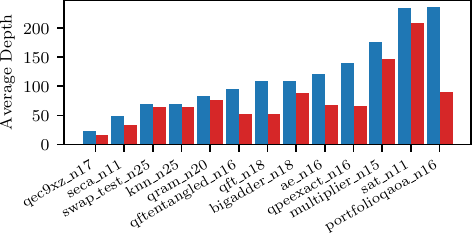}
        \caption{Critical path depth for Square-Lattice topology}
        \label{fig:square-depth}
    \end{subfigure}
    \begin{subfigure}[b]{\columnwidth}
        \centering
        \includegraphics[width=\columnwidth]{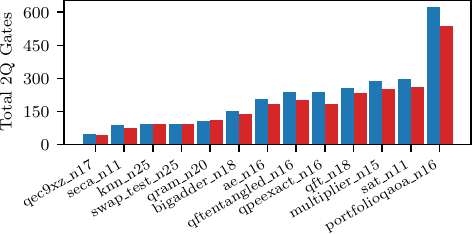}
        \caption{Total gate cost for Square-Lattice topology}
        \label{fig:square-total}
    \end{subfigure}
    \caption{Optimization results on two topologies: comparing between Qiskit-SABRE and MIRAGE.}
    \label{fig:results}
\end{figure*}

We evaluated MIRAGE targeting two production quantum machine topologies: the 57Q Heavy-Hex and the 6x6 Square-Lattice, tracking three metrics: critical path depth, total gate cost, and the number of SWAP gates.  We report both an average of the improvements as well as a weighted average improvement based on the circuit size.  The results of MIRAGE on quantum circuit routing are shown in Figure~\ref{fig:results}. 

% % Results for w5 Heavy-Hex
% depth': {'average_change': -31.18925974200507,
%   'aggregrate_change': -33.40650586991774,
%   'best_circuit': 'qpeexact_n16',  0.948 \% sub rate
%   'worst_circuit': 'knn_n25'},  0.425\% sub rate
% total': {'average_change': -16.96877185481387,
%   'aggregrate_change': -22.35381939803622,
%   'best_circuit': 'qpeexact_n16', 0.948 \% sub rate
%   'worst_circuit': 'qram_n20'},  0.309  \% sub rate
% swaps': {'average_change': -56.189143303765505,
%   'aggregrate_change': -75.8151733091156,
%   'best_circuit': 'portfolioqaoa_n16',  0.987\% sub rate
%   'worst_circuit': 'qram_n20'}, 0.309  \% sub rate

\textbf{For the Heavy-Hex topology, we observed an average and weighted average decrease of 31.19\% and 33.41\%, respectively, in circuit depth} (Fig.~\ref{fig:hex-depth}). The largest decrease in depth was `qpeexact\_n16' with a mirror gate acceptance rate of 0.948\%, while the least was `knn\_n25' with a mirror gate acceptance rate of only 0.425\%. \textbf{In terms of total gate costs, the average decrease was 16.97\%, for an aggregrate decrease of 22.35\%} (Fig.~\ref{fig:hex-total}). The circuit with the greatest reduction of total gates was `qpeexact\_n16`, while the least was `qram\_n20' with a mirror gate acceptance rate of only 0.309\%. \textbf{The average decrease of \sw{}s was 56.19\%, or a weighted average decrease of 75.82\%.} The greatest decrease in \sw{} gates was `portfolioqaoa\_n16' with a mirror gate acceptance rate of 0.987\%, while the worst was`qram\_n20`.

% % Results for the 6x6 topology
% depth': {'average_change': -29.58293759547691,
%   'aggregrate_change': -32.08691273036692,
%   'best_circuit': 'portfolioqaoa_n16', 100\% sub rate
%   'worst_circuit': 'swap_test_n25'},  0.344\% sub rate
% total': {'average_change': -10.253081206893013,
%   'aggregrate_change': -12.342163085432158,
%   'best_circuit': 'qpeexact_n16',  0.933 \% sub rate
%   'worst_circuit': 'qram_n20'},  0.353   \% sub rate
% swaps': {'average_change': -59.862211355192535,
%   'aggregrate_change': -77.61235151089284,
%   'best_circuit': 'seca_n11', 0.282 \% sub rate
%   'worst_circuit': 'qram_n20'},  0.353   \% sub rate

\textbf{For the Square-Lattice topology, we observed an average and weighted average decrease of 29.58\% and 32.09\%, respectively in circuit depth} (Fig.\ref{fig:square-depth}). The largest decrease in depth was 'portfolioqaoa\_n16' with a mirror gate acceptance rate of 100\%, while the least was 'swap\_test\_n25' with a mirror gate acceptance rate of only 0.344\%. \textbf{In terms of total gate costs, the average decrease was 10.25\%, for a weighted average decrease of 12.34\% }(Fig.\ref{fig:square-total}). The circuit with the greatest reduction of total gates was 'qpeexact\_n16', while the least was 'qram\_n20' with a mirror gate acceptance rate of only 0.353\%. \textbf{The average decrease of SWAP gates was 59.86\%, or a weighted average decrease of 77.61\%}. The greatest decrease in SWAP gates was 'seca\_n11' with a mirror gate acceptance rate of 0.282\%, while the worst was again 'qram\_n20'.

\subsection{Implementation Improvements: Caching and Parallelism}

% Motivate: we added things that will make it much more expensive
To ensure a reasonable runtime of MIRAGE we profiled the tool and address the computational bottlenecks associated with the intermediate representation of the DAG nodes using coordinates. Profiling revealed that the conversion of a unitary to a coordinate and the \texttt{UnitaryGate} constructor were among the most expensive operations. To mitigate these issues, we introduced caching and parallelism into our approach.

% mirror gates in the intermediate layer
In the intermediate layer, when a mirror gate is accepted, we replace the entire DAG node with a new unitary, rather than appending a gate that would disrupt consolidation. To efficiently build this unitary, we removed the costly calls to \texttt{is\_unitary} and \texttt{is\_identity} in the \texttt{UnitaryGate} constructor, as we know that mirroring will always preserve unitarity. Finally, instead of computing the new mirror gate's coordinate, we use an equation to convert the original coordinate to the mirror coordinate (Eq.~\ref{eq:mirror-coordinates}). Every \sw{} gate added by MIRAGE comes with the \sw{} coordinate manually annotated to the node.

\begin{figure}[tbp]
    \centering
    \begin{subfigure}[b]{\columnwidth}
        \centering
        \includegraphics[width=0.9\columnwidth]{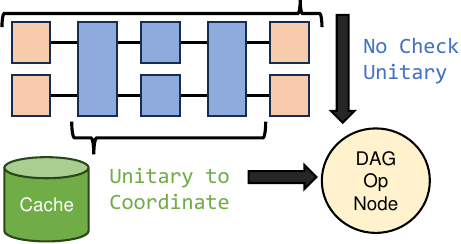}
        \caption{Caching mechanism for building unitary Weyl coordinates}
        \label{fig:cache-diagram}
    \end{subfigure}
    
    \begin{subfigure}[b]{\columnwidth}
        \includegraphics[width=0.9\columnwidth]{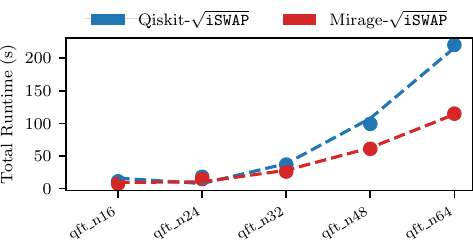}
        \caption{Measured runtimes of Quantum Fourier Transform}
        \label{fig:qft-runtimes}
    \end{subfigure}
    \caption{MIRAGE performance enhancements}
    \label{fig:cache}
\end{figure}

% ConsolidateBlocks
To further speed up conversion from unitary to coordinates,rewrite the \texttt{ConsolidateBlocksocks} pass. Rather than multiplying all the operators into a new unitary, we first skip the exterior 1Q gates (since these will not change the coordinate), use that unitary as the cache key, annotate the block, and then multiply in the remaining gates (Figure~\ref{fig:cache}). This approach increases cache hits by removing the 1Q unitaries that would otherwise lead to different unitaries and ensures that each unitary has an annotated coordinate, eliminating the need for repeated calculations.

% polytope cache
Finally, we implemented an LRU software cache (lookup table) for each circuit polytope, significantly reducing the time spent querying the same coordinates. This cache ensures that each coordinate only needs to be queried once, thus reducing the number of expensive iterations through polytopes and calls to the \texttt{has\_element} function.

% DATA, how much faster are all these changes?
Our method scales better than Qiskit's Python transpiler. \textbf{For a 64Q QFT circuit, MIRAGE ran 47.9\% faster than Qiskit.} The speed of the code determines the number of independent trials, which in turn determines the quality of solution from the transpiler.  

Thus, MIRAGE can effectively improve circuit depth on different circuit topologies just through efficient utilization of mirror gates without increasing transpilation runtime.

\section{Related Work}
% important previous work on the cns gate
% this is too relevant to remove, but was not included in the HPCA first draft (I think we should add it back in)
Previous works have identified \cns{} optimizations in some common circuit elements, \textit{e.g.}, the Toffoli and Fredkin 3Q gates can be decomposed using \cnot{} and \cns{} gates allowing for optimization at the gate-decomposition level~\cite{schuch2003natural, liu2023qcontext, cruz2023shallow}. The \cns{} gate can also be found in stabilizer measurements of error-correcting codes~\cite{schuch2003natural, antipov2023realizing, simakov2022scalable} in entanglement purification protofcols~\cite{tanamoto2008efficient}, and QAOA circuits~\cite{ji2023optimizing, tan2021optimal, hashim2021optimized}. In QFT circuits, the fractional controlled-phase gates can be replaced by \cns{} gates, allowing the QFT circuit to be implemented solely using \iswap{} gates~\cite{wang2011efficient}.  Moreover, the Diamond Gate, a native 4Q gate, can be recast into \cns{} gates, which can be used to build controlled-phase operations~\cite{bahnsen2022application}. 

Additionally, several modifications have been proposed to further improve the effectiveness of SABRE.  The most relevent are ``Not all Swaps have the Same Cost'' (NASSC), which refines the SABRE \sw{} selection heuristic to select the most efficient \sw{} by additionally considering potential \cnot{} gate cancellations due to encoded commutation rules~\cite{liu2022not}.  The transpiler analyzes the set of predecessor and successor nodes for each \sw{} candidate to identify potential cancellations.  For example, it considers a \sw{} to be three inverted \cnot{}s and it looks for cases where those \cnot{}s can be commuted in the circuit such that the first and last \cnot{} from two \sw{} gates can be cancelled.  Like MIRAGE, NASSC does consider decomposition, but it is a very specific optimization to \cnot{}, while MIRAGE is much more general to the basis gate.

Permutation-aware Synthesis and Permutation-aware Mapping (PAS+PAM) searches for alternative decompositions of unitary blocks before committing any particular node to the mapped DAG~\cite{liu2023tackling}. PAS+PAM considers the potential for synthesizing an equivalent operation with permuted input and output qubits, which may reduce routing overhead. This is similar to MIRAGE.  However, PAS+PAM, like NASSC, focuses on \cnot{} basis gates whereas MIRAGE is general to any basis gate.  While PAS+PAM attempts to fully link decomposition and routing, MIRAGE uses decomposition information determined through monodromy polytopes, again which is agnostic to the type of basis gate, to track circuit depth and promote gate combination while routing to maintain an efficient runtime rather than fully combining the steps.  Moreover, MIRAGE reports $>$30\% reduction in circuit depth versus 18\% reported by PAS+PAM to the same Qiskit baseline.

\section{Conclusion}
In this work, we have proposed MIRAGE, a strategy that integrates routing and decomposition in the transpiler, breaking the traditional abstraction of routing and decomposition. %This approach has shown to be effective in reducing the depth and total gate cost of quantum circuits.In \emph{MIRAGE}, we presented a novel approach to circuit transpilation for \sqiswap{} native hardware. Our method integrates decomposition cost considerations when routing \sw{} operations, simplifying circuit depth and demonstrating the utility of mirror gates. 
Our experiments show tangible improvements: For the Heavy-Hex topology, there was an average reduction of 31.19\% in circuit depth, 16.97\% decrease in total gate count, accomplished by a noteworthy 56.19\% decrease of \sw{}s. Meanwhile, the Square-Lattice topology experienced an average decrease of 29.58\% in circuit depth and a 59.86\% reduction in \sw{} gates. These results emphasize the value of our approach and the importance of aligning circuit design with topology. The insights from \emph{MIRAGE} offer a foundation for refining transpilation strategies, potentially benefiting a wider range of topologies and circuit designs.  

Future work aims to integrate MIRAGE directly into the transpiler pipeline using plugins%, enhancing MIRAGE's efficiency and versatility
.
This allows us to select the routing trial that minimizes the estimated circuit depth, making each iteration of our program more geared towards depth reduction.  Lastly, finding approximate decompositions without Monte Carlo methods could be done using quadratic programming or affine subspace projections to accelerate our approach for other basis gates.

\section{Software Availability}
MIRAGE is open source and available online through \url{https://github.com/Pitt-JonesLab/mirror-gates} and will be provided for artifact evaluation.

\section*{Acknowledgements}
This work is partially supported by The Charles E. Kaufman Foundation of The Pittsburgh Foundation under New Initiative Award KA2022-129519 and the University of Pittsburgh via a SEEDER grant.

%%%%%%% -- PAPER CONTENT ENDS -- %%%%%%%%

%%%%%%%%% -- BIB STYLE AND FILE -- %%%%%%%%
\bibliographystyle{IEEEtran}
\bibliography{refs}
%%%%%%%%%%%%%%%%%%%%%%%%%%%%%%%%%%%%

\end{document}